\documentclass[11pt,a4paper]{article}
\usepackage{geometry}
\usepackage{authblk}
\usepackage{amsmath}
\usepackage{hyperref}
\usepackage{cite}

\begin{document}

\title{Physical Problems Admitting Heun-to-Hypergeometric Reduction}

\author{Pelin Ayd{\i}ner}
\author{Tolga Birkandan\footnote{E-mail address: birkandant@itu.edu.tr}}

\affil{Istanbul Technical University, Department of Physics, \\
Istanbul, Turkey}

\maketitle

\begin{abstract}
The Heun's equation with its four regular singularities emerges in many applications in science. Despite the growing interest of the scientific community, the literature has many gaps in conceptual mathematical aspects of this equation. Moreover, the translation of the mathematical language for non-mathematicians in making contemporary ideas applied in physical research is not also well developed. In this paper, Maier's Heun-to-hypergeometric reduction cases are studied in detail for four problems in quantum mechanics: (1) The Schrodinger's equation for the Coulomb problem on a 3-sphere, (2) the s-wave bound state equation in the problem of the attractive inverse square potential, (3) the equation for the limit density function for the discrete-time quantum walk and (4) charged particle under magnetic field and coulomb force. All these problems give rise to the Heun's equation in their mathematical formulation. A Sage code is given in the appendix for the calculation of some Heun identities.
\end{abstract}



\newpage

\tableofcontents
\section{Introduction}
As new physical solutions giving the Heun's function emerge frequently, the need for understanding the mathematical aspects of the Heun's equation becomes more important \cite{hortacsu}. Although the number of occurrences of this equation in modern applications is increasing, the number of studies related to the mathematical analysis of the Heun's equation is very few. Among these limited number of mathematical works, articles on reducing the Heun's equation to the hypergeometric equations have particular importance since establishing a relationship with well-studied equations in the literature gives a deeper insight about the less known Heun's equation and its confluent types \cite{kuiken,vidunas,vidu2,take,sokho,gaspar,houn,hoej,mahu,jaick,ishk1,cheb,ishk2,ishk3,ishk4,ishk5,maierheuntohyper}. Among very few articles that refer to the physical applications of these reductions we can cite Kwon et al. \cite{kwon} and Cunha and Christiansen's \cite{cunha,cunha2} works.

In this paper, we will give examples of Heun-to-hypergeometric reductions. The first one is the quantum mechanical Coulomb problem on a 3-sphere analyzed by Bellucci et al. \cite{bellucci}. They studied the Schrodinger's equation in the generalized parabolic coordinates and found that the equation could be written in the form of the Heun's general equation. The second problem is the s-wave bound state equation in the attractive inverse square potential studied by Bouaziz and Bawin \cite{bouaziz}. The third example is the equation for the limit density function for the discrete-time quantum walk as studied by Konno et al. where they found a relationship between the limit density function and the Heun's equation \cite{konno}. The last example is Ralko and Truong's calculation on the motion of the charged particles on a sphere under constant magnetic force \cite{ralko}. For these problems, we will study all possible reduction cases.

In the second section, the identities and the limiting cases which are needed for the Heun-to-hypergeometric reductions given by Maier will be summarized. The subsequent sections contain the examples in which the limiting cases are studied in detail. In the appendix, a symbolic Sage code is given for the calculation of some Heun identities that will be covered in the second section \cite{sage}.
\section{Heun's equation and its reductions}
The general Heun's equation is given by
\begin{equation}
H^{\prime \prime }+\left( \frac{\Gamma }{z}+\frac{\Delta }{z-1}+\frac{\epsilon }{z-d}\right) H^{\prime }+\frac{abz-q}{z(z-1)(z-d)}H=0, \label{genheun}
\end{equation}
where the prime denotes differentiation with respect to $z$ and the relation $a+b+1=\Gamma +\Delta +\epsilon $ should hold. The equation has four regular singularities at $\{0,1,d,\infty \},$ ($d\neq 0,1$). The solution of this equation is the Heun's function and it is denoted by $H(d,q;a,b,\Gamma,\Delta ;z)$ \cite{ronveaux,slavyanov}.

According to Maier \cite{maierheuntohyper}, there are mainly five cases in which this equation reduces to a hypergeometric equation:

1. $\epsilon =0$ and $q=abd$

2. $\Delta =0$ and $q=ab$

3. $\Gamma =0$ and $q=0$

4. $ab=0$ and $q=0$

5. Nontrivial reductions

The fourth case implies the absence of the singularity at infinity as given by the Definition (2.1) of Maier \cite{maierheuntohyper} and it is called the ``trivial" case for the Heun's equation.

One can study the nontrivial reductions when the singular points satisfy the harmonic (equally spaced collinear points) or equianharmonic (equilateral triangle) conditions as given by the Theorem (3.1) of Maier \cite{maierheuntohyper}. We will restrict our interest to the cases exhibiting harmonic behavior among the singular points (i.e. the real case). We will use the two identities given by Maier to change the $d$ values to $\frac{d}{d-1}$ or $\frac{1}{d}$ \cite{maier192}.

The first identity that could be used was given in Table 2 (line 5) of Maier's article which contains the 192 solutions of the Heun equation \cite{maier192}, namely,
\begin{equation}
H(d,q;a,b,\Gamma ,\Delta ;z)=(1-z)^{-a}H\left(\frac{d}{d-1},\frac{ad\Gamma -q}{d-1};a,a-\Delta +1,\Gamma ,a-b+1;\frac{z}{z-1}\right),
\end{equation}
and the second one is given in Table 2 (line 9) of the same article, namely,
\begin{equation}
H(d,q;a,b,\Gamma ,\Delta ;z)=H\left( \frac{1}{d},\frac{q}{d};a,b,\Gamma,a+b-\Gamma -\Delta +1;\frac{z}{d}\right) .
\end{equation}
A short Sage code is given in the appendix to facilitate the application of these identities easily \cite{sage}. The user can add other identities in the code using similar definitions. These identities enable us to interchange the singularity in the set $\left\{ d,\frac{d}{d-1},\frac{1}{d}\right\} $.

According to the Theorem (3.7) of Maier \cite{maierheuntohyper}, the Heun's equation is nontrivial if $ab\neq 0$ and $q\neq 0$ and the solution can be reduced to a hypergeometric function by a formula of the type $H(t)=F(R(t))$, with a rational function $R$ only if the parameters satisfy $q=abp$ with $(d,p)$ equal to one of the 23 pairs stated in the theorem with a given degree for $R(t)$. For example, by using the identities stated above we can have $d=\{-1,2,\frac{1}{2}\}$ which corresponds to a harmonic cross ratio orbit according to Theorem 3.1 (1a) of Maier \cite{maierheuntohyper}. The related pairs are $(d,p)=(-1,0)$, $(\frac{1}{2},\frac{1}{2})$, $(2,1)$ and they come with degree $2$ or $4$ for the function $R$.
\section{Quantum mechanical Coulomb problem on a 3-sphere}
Bellucci and Yeghikyan studied the quantum mechanical Coulomb problem on a 3-sphere and found that the resulting equation could be written in the form of the general Heun's equation. This equation reads
\begin{equation}
H^{\prime \prime }+\left( \frac{\Gamma }{z}+\frac{\Delta }{z-1}+\frac{\epsilon }{z+1}\right) H^{\prime }+\frac{abz-q}{z(z-1)(z+1)}H=0,
\end{equation}
where
\begin{eqnarray}
\Gamma  &=&1-\sqrt{1+E+i\gamma },\Delta =\epsilon =|m|+1, \\
q &=&i\frac{\beta }{2}, \\
a &=&1+|m|+\frac{\sqrt{1+E-i\gamma }-\sqrt{1+E+i\gamma }}{2}, \\
b &=&1+|m|-\frac{\sqrt{1+E-i\gamma }+\sqrt{1+E+i\gamma }}{2}.
\end{eqnarray}
Here, $\beta $ is a separation constant, $m$ is an natural number associated with the axial symmetry, $\gamma $ is a parameter defined in the suggested solution and $E$ is the energy. We have the third regular singular point at $d=-1$ in our case. From now on, we will use $m$ instead of its absolute value $|m|$ \cite{bellucci}.

The cases with \{$\epsilon =0$ and $q=abd$\} and \{$\Delta =0$ and $q=ab$\} are inapplicable for our problem as $\Delta =\epsilon =m+1$ and this value is restricted to nonzero values of $m$ which is actually the absolute value of this parameter. Now let us study the special cases where the equation reduces to a hypergeometric equation.
\subsection{The case with $\Gamma =0$ and $q=0$}
When $\Gamma \ $and $q$ are zero, the parameters yield $E=-i\gamma $ and $\beta =0$. This reduces our equation to the Legendre equation with three regular singularities at $\{-1,1,\infty \}$ as
\begin{equation}
H=(z^{2}-1)^{-m/2}\left[ P_{\lambda }^{m}\left( z\right) +Q_{\lambda}^{m}\left( z\right) \right] ,
\end{equation}
where $\lambda =\frac{\sqrt{1-2i\gamma }-1}{2}$. This case is merely of mathematical interest because of the imaginary energy condition.
\subsection{The trivial case\ $\{ab=0$ and $q=0\}$}
When the product of $a$ and $b$ is zero, the energy values are restricted to
\begin{equation}
E=m(m+2)-\frac{\gamma ^{2}}{4(1+m)^{2}},
\end{equation}
and we set $\beta =0$ to satisfy $q=0$. This energy condition has a physical importance as the energy spectrum found by the power series expansion of the Heun's equation was given by
\begin{equation}
E_{n}=(n+m)(n+m+2)-\frac{\gamma ^{2}}{4(n+1+m)^{2}},  \label{enbellu}
\end{equation}
in the original article \cite{bellucci} and our result gives the ground state ($n=0$) relating with this energy spectrum. The quantum number $n$ was also related to the parameter $q$ in the article which is zero in our case. Moreover, the general problem has been solved in terms of the Heun's polynomials in the original article and our reduction shows that the solution can be stated by the hypergeometric function for the ground state
\begin{equation}
H=F\left( \frac{m+1}{2}-\frac{\sqrt{C_{1}}-\sqrt{C_{1}-8i\gamma }}{8},\frac{m+1}{2}-\frac{\sqrt{C_{1}}+\sqrt{C_{1}-8i\gamma }}{8};1-\frac{\sqrt{C_{1}}}{4};z^{2}\right),
\end{equation}
where
\begin{equation}
C_{1}=4(E+i\gamma +1),
\end{equation}
noticing that $\sqrt{1+E+i\gamma }$ factor also appears in the original parameters. This result shows a physical correspondence of the Heun-to-hypergeometric reduction.
\subsection{Nontrivial reductions}
The easiest approach would be using the equation (3.8) of Maier \cite{maierheuntohyper} with $(d,p)=(-1,0)$, namely
\begin{equation}
H\left( -1,0;a,b,\Gamma ,\frac{a+b-\Gamma +1}{2};z\right) =F\left( \frac{a}{2},\frac{b}{2};\frac{\Gamma +1}{2};z^{2}\right) , \label{minusone}
\end{equation}
as $d=-1$ originally. The condition $\Delta =\frac{a+b-\Gamma +1}{2}=m+1$ is automatically satisfied using the physical parameters. Therefore, in this case, the only constraint we have is $q=0$ with $\beta =0$ as we need to have $q=abp$. We cannot deduce an energy condition from the parameter constraints for this case. $q$ being zero however, implies that the case is also related with the ground state. The relation between the quantum states and the parameter $q$ will be clear in the next case.

We can also use $(d,p)=(2,1)$ after appyling the identities of the Heun's function. The third singular point is $d=-1$ but we can change it to $d=2$ using the identity given in Table 2 (line 17) of Maier's article of the 192 solutions \cite{maier192}. It reads
\begin{equation}
H(-1,q;a,b,\Gamma ,\Delta ;z)=(1+z)^{-a}H\left( 2,a\Gamma -q;a,-b+\Delta+\Gamma ,\Gamma ,\Delta ;\frac{2z}{z-1}\right) ,
\end{equation}
and $(d,p)=(2,1)$ yields   $a\Gamma -q=a(\Delta +\Gamma -b)$ by the condition $q=abp$. Using this, we can specify the value of the corresponding energy as follows
\begin{equation}
E_{n_{reduc}}={\frac{-\beta ^{4}+4\gamma \beta ^{3}-2\left[ 3\gamma ^{2}+2({m+}1)^{2}\right] \beta ^{2}+4\left[ \gamma ^{3}+2({m+}1)^{2}\gamma \right]\beta -\gamma ^{4}+4m\left( {m+2}\right) (m+1)^{2}\gamma ^{2}}{4\left(m+1\right) ^{2}\left( \beta -\gamma \right) ^{2}}.}
\end{equation}
The above energy value coincides with the one we found in the ``trivial" case for $q=0$ (i.e. $\beta =0$). We have a nonzero $q$ parameter in this case, therefore using this energy condition, we may find the counterpart of the quantization parameter $n$ associated with the reduction case in terms of our physical parameters. This parameter should also satisfy $n_{reduc}=0$ if $\beta =0$. Then we can calculate,
\begin{equation}
n_{reduc}=-\frac{(1+m)\beta }{\beta -\gamma },
\end{equation}
as the quantization parameter associated with our reduction case.

We can reduce the Heun's function using the equation (3.5c) of Maier \cite{maierheuntohyper}, namely
\begin{equation}
H(2,ab;a,b,\frac{a+b+2}{4},\frac{a+b}{2};t)=F\left( \frac{a}{4},\frac{b}{4};\frac{a+b+2}{4};1-4\left[ t\left( 2-t\right) -\frac{1}{2}\right] ^{2}\right).  \label{maier35a}
\end{equation}
Matching the two Heun functions
\begin{equation}
H\left( 2,a\Gamma -q;a,-b+\Delta +\Gamma ,\Gamma ,\Delta ;\frac{2z}{z-1}\right) =H\left(2,ab;a,b,\frac{a+b+2}{4},\frac{a+b}{2};t\right),
\end{equation}
we obtain
\begin{equation}
n_{reduc}=\frac{m}{2}-1,
\end{equation}
by choosing the appropriate \{$\beta ,\gamma $\} values emerging from the matching conditions. Remembering $m$ being a positive integer, we can deduce that the case restricts the $m$ values to only even integer numbers. We finally write
\begin{equation}
H=F\left( \frac{a}{4},\frac{-b+\Delta +\Gamma }{4};\frac{a-b+\Delta +\Gamma+2}{4};1-4\left[ -\frac{z^{2}+6z+1}{2(z-1)^{2}}\right] ^{2}\right) ,
\end{equation}
as the solution reduced to the hypergeometric function.
\section{s-wave bound state equation in the problem of the attractive inverse square potential}
The equation which Bouaziz and Bawin obtained for the wave function in their article can be rewritten as
\begin{equation}
H^{\prime \prime }+\left( \frac{\Gamma }{z}+\frac{\Delta }{z-1}+\frac{\epsilon }{z-d}\right) H^{\prime }+\frac{abz-q}{z(z-1)(z-d)}H=0,
\end{equation}
where
\begin{eqnarray}
\Gamma &=&\frac{3}{2},\Delta =\frac{1}{2}-\omega _{4},\epsilon =2, \\
q &=&\frac{3}{2}+\frac{\kappa }{1-2\omega }, \\
a &=&\frac{1}{2}\left( 3-\omega _{4}-\nu \right) ,b=\frac{1}{2}\left(
3-\omega _{4}+\nu \right) , \\
\nu &=&\sqrt{(\omega _{4}-1)^{2}-\frac{4\kappa }{1-2\omega }}, \\
d &=&\frac{2\omega }{2\omega -1}.
\end{eqnarray}
Note that we changed the sign of the accessory parameter $q$ in the original paper to the standard form of the Heun's equation we used. Here, $\omega _{4}$ and $\kappa $ depend on $\beta $'s coming from the modified (generalized) commutation relations between the position and momentum operators, $[\hat{X}_{i},\hat{P}_{j}]=i\hbar \left[ \left( 1+\beta \hat{P}^{2}\right) \delta _{ij}+\beta ^{\prime }\hat{P}_{i}\hat{P}_{j}\right] $ and ($\beta, \beta ^{\prime })>0$. The relations can be given as $\omega_{4}=\frac{\beta }{\beta +\beta ^{\prime }}$ and $\kappa=\frac{m\alpha }{2\hbar }$ where $\alpha =\frac{\gamma -\beta ^{\prime }\left( \frac{D-1}{2}\right) }{\beta +\beta ^{\prime }}>0$ and $\gamma $ is an arbitrary constant \cite{bouaziz}.

The case with \{$\epsilon =0$ and $q=abd$\} is inapplicable as the parameter $\epsilon $ is fixed as $\epsilon =2$. The case with \{$\Gamma =0$ and $q=0$\} is also not possible since we have $\Gamma =\frac{3}{2}$.
\subsection{The case with $\Delta =0$ and $q=ab$}
$\Delta =0$ yields $\omega _{4}=\frac{1}{2}$ and also $q=ab$ yields the same condition. Therefore $\omega _{4}=\frac{1}{2}$ is the sufficient condition to have this case to transform from Heun-to-hypergeometric equation as
\begin{equation}
H=F\left( \frac{5\,}{4}-\frac{\sqrt{2\,\omega -1+16\,\kappa }}{4\sqrt{2\,\omega -1}},\frac{5\,}{4}+\frac{\sqrt{2\,\omega -1+16\,\kappa }}{4\sqrt{2\,\omega -1}},\frac{3}{2},\frac{(2\omega -1)}{2\omega}z\right) .
\end{equation}
The result coincides with the special case $\beta =\beta ^{\prime}$ that the authors have studied and calculated the corresponding energy spectrum in their article\cite{bouaziz}. The result is not unexpected as setting $\omega _{4}=\frac{1}{2}$ directly in the definition $\omega _{4}=\frac{\beta }{\beta +\beta ^{\prime }}$ gives $\beta =\beta^{\prime }$ which was our only constraint in this reduction case.
\subsection{The trivial case\ $\{ab=0$ and $q=0\}$}
$q=0$ yields $\kappa =3\omega -\frac{3}{2}$ and $ab=0$ yields $%
\omega _{4}=\frac{1}{2}$. Using the previous result we obtain
\begin{equation}
H=F\left( 0,\frac{5\,}{2},\frac{3}{2},\frac{(2\omega -1)}{2\omega }z\right)=1,
\end{equation}
which is not interesting physically.
\subsection{Nontrivial reductions}
As we have $d=\frac{2\omega }{2\omega -1}$, we may seem to be free to choose a $(d,p)$ pair among the 23 pairs stated in the Theorem (3.7) of Maier \cite{maierheuntohyper}. Picking $(d,p)=(2,1)$ results in $\omega_{4}=\frac{1}{2}$, which was also found in the case with \{$\Delta =0$ and $q=ab$\}. This result makes this case trivial. The pair $(d,p)=(-1,0)$ resulting in $\omega =\frac{1}{4}$ and $\kappa =-\frac{3}{4}$ is also prohibited as $\kappa $ is restricted to have positive values ($\alpha >0$). In general, the pairs with $p=1$ yields the case $\omega _{4}=\frac{1}{2}$ \ which is trivial and the pairs with $p=0$ results in negative $\kappa $ values for the permitted $d$ values, namely $d\in $\{$-1,-3,-\frac{1}{3}$\}. The other pairs also result in nonpositive $\kappa $'s as $\omega _{4}$ should also be positive. The $\kappa $ values in these cases depend on $\omega _{4}$'s such that we always need a negative $\omega _{4}$ to have a positive $\kappa $ value. Therefore, we can conclude that the problem cannot be reduced nontrivially via harmonic singular points.
\section{The equation for the limit density function for the discrete-time quantum walk}
The relationship between the limit density function and the Heun's equation found by Konno et al. can be stated as
\begin{equation}
H^{\prime \prime }+\left( \frac{\Gamma }{z}+\frac{\Delta }{z-1}+\frac{\epsilon }{z-d}\right) H^{\prime }+\frac{abz-q}{z(z-1)(z-d)}H=0,
\end{equation}
where
\begin{eqnarray}
\Gamma &=&\frac{1}{2},\Delta =2,\epsilon =\frac{3}{2},a=b=\frac{3}{2}, \\
q &=&\frac{2d+1}{4},
\end{eqnarray}
and $d>0$ \cite{konno}. The solution of this equation can be given in terms of the Heun's function but in this case it can be also written as
\begin{equation}
H=\frac{\sqrt{1-d}}{\pi (1-z)\sqrt{d-z}},  \label{konnosol}
\end{equation}
and this function is still a solution of the equation without the factor $\pi$ and $(z-1)$ instead of $(1-z)$ in the denominator or any numerical value in the numerator instead of $\sqrt{1-d}$. This solution corresponds to the limit density function for the discrete-time quantum walk. Therefore, the choice of the parameters in this solution is physical.

We do not have the first four reduction cases as the parameters have fixed values and they do not satisfy the conditions needed for these cases.

For the non-trivial case, we will do the following calculation:

The singular point $d$ is restricted to have positive values. Therefore we exclude the $(d,p)$ pairs with negative $d$ values. Another restriction is that the $a$ and $b$ parameters have fixed values to give $2d+1=9p$ using $q=abp$. Only the pair $(d,p)=(4,1)$ satisfies this condition. We have
\begin{equation}
H\left(4,ab;a,b,\frac{1}{2},\frac{2(a+b)}{3};z\right)=F\left[ \frac{a}{3},\frac{b}{3};\frac{1}{2};1-(z-1)^{2}\left( 1-\frac{z}{4}\right) \right] ,
\end{equation}
as the equation (3.5b) in \cite{maierheuntohyper} and thus we have
\begin{equation}
H\left(4,\frac{9}{4};\frac{3}{2},\frac{3}{2},\frac{1}{2},2;z\right)=F\left[ \frac{1}{2},\frac{1}{2};\frac{1}{2};1-(z-1)^{2}\left( 1-\frac{z}{4}\right) \right] ,
\end{equation}
as the reduction to the hypergeometric function. We should note that the pair $(d,p)=$ $(4,1)$ corresponds to degree $3$ for the function $R$ as can be seen from the result. We can reduce our result further
\begin{equation}
F\left[ \frac{1}{2},\frac{1}{2};\frac{1}{2};1-(z-1)^{2}\left( 1-\frac{z}{4}\right) \right] =\frac{2}{(z-1)\sqrt{4-z}},
\end{equation}
as it has the appropriate parameters \cite{abraste}.

We can conclude that we recovered a mathematically similar limit density function for a special case in which the solution could be reduced to the hypergeometric function, but with different parameters. These parameters should be fixed using the physical problem. We can write the hypergeometric equation corresponding to our special case and easily see that the original solution given by the equation (\ref{konnosol}) satisfies the hypergeometric equation for $d=4$.
\section{Charged particle under magnetic field and Coulomb force}
The studies of Ralko and Truong \cite{ralko} on the motion of the charged particles on a sphere under constant magnetic force and competing forces propounded that the equation of the system can be stated in the form of the Heun's equation (\ref{genheun}) with the parameters
\begin{gather}
\Gamma =2a^{\prime }+1, \Delta =\varepsilon =b^{\prime }+1,
d =-1, q =-\frac{4R}{l_{0}}, \\
ab =(a^{\prime }+b^{\prime })(a^{\prime }+b^{\prime }+2)-4(\varepsilon^{\prime }+S^{2}), \label{abralko}
\end{gather}
and
\begin{gather}
a^{\prime } =|S-m|, b^{\prime } =|S+m|, \\
l_{0} =\sqrt{\frac{\hbar ^{2}}{MKe^{2}}}, S =\frac{R^{2}}{l_{B}^{2}}, l_{B}^{2} =\sqrt{\frac{\hbar }{eB}}.
\end{gather}
Here, $l_{0}$ is the Bohr radius, $m$ is an integer related to the spherical harmonics, $S$ is the strength of the magnetic monopole generating the magnetic field $B$, $l_{B}$ is a magnitude defined as the magnetic length, $R $ is the radius of the sphere and $\varepsilon ^{\prime }$ is the quantized energy levels.

The first three cases of the given reductions can not be applied in this problem. The restrictions of $\varepsilon=0$, $\Delta=0$ and $\Gamma=0$ results in negative $a^{\prime}$ and $b^{\prime}$ values which are not allowed.
\subsection{The case with $ab =0$ and $q=0$}
The condition $ab=0$ yields
\begin{equation}
(a^{\prime }+b^{\prime })(a^{\prime }+b^{\prime }+2)-4(\varepsilon ^{\prime}+S^{2})=0.
\end{equation}
We can obtain the energy formula as
\begin{equation}
\varepsilon ^{\prime }=\frac{(a^{\prime }+b^{\prime })(a^{\prime }+b^{\prime}+2)}{4}-S^{2},
\end{equation}%
which is exactly the same energy formula stated in the article \cite{ralko} in the part where the limit cases are considered to determine an energy condition. We can have this limit naturally via $q=-\frac{4R}{l_{0}}=0$ which implies that the Bohr radius $l_{0}$ goes to infinity and that the Coulomb repulsion vanishes. In this case, we are able to reduce the solution to a integral, namely
\begin{equation}
H\left( z\right) =C_{1}+C_{2}\int \!{z}^{-\Gamma }\left( z-1\right)^{-\Delta }\left( z-d\right) ^{-\epsilon }\,dz,
\end{equation}
where $C_{1}$ and $C_{2}$ are some constants.
\subsection{Nontrivial reduction}
We can consider the case with $(d,p)=(-1,0)$ with the reduction given by the equation (\ref{minusone}). We need to restrict $q$ using $q=abp$ as $q=0$. Thus, we need the case where the Coulomb repulsion vanishes ($l_{0}\rightarrow \infty $) to have $q=-\frac{4R}{l_{0}}=0$. We do not have $a$ and $b$ parameters individually in the original parametrization, but we have their multiplication as in the equation (\ref{abralko}) and we know that the original Heun's equation needs to satisfy $a+b+1=\Gamma+\Delta +\varepsilon$. Using these two conditions we obtain
\begin{gather}
a =a^{\prime }+b^{\prime }+1+\sqrt{4\,{S}^{2}+4\,\varepsilon ^{\prime }+1},\\
b =a^{\prime }+b^{\prime }+1-\sqrt{4\,{S}^{2}+4\,\varepsilon ^{\prime }+1},
\end{gather}
and we can use them interchangeably ($a\leftrightarrow b$). In order to have this nontrivial reduction, we need to have
\begin{equation}
\Delta =\frac{a+b-\Gamma +1}{2}=b^{\prime }+1.
\end{equation}
Using our parameters, we see that $\Delta =b^{\prime }+1$ is satisfied. Therefore we can write the reduction as
%
\begin{equation}
H=F\bigg( \frac{a^{\prime }+b^{\prime }+1+\sqrt{4\,{S}^{2}+4\,\varepsilon^{\prime }+1}}{2}
,\frac{a^{\prime }+b^{\prime }+1-\sqrt{4\,{S}^{2}+4\,\varepsilon ^{\prime }+1}}{2};a^{\prime }+1;z^{2}\bigg),
\end{equation}
for this problem, under the given conditions.
\section{Conclusion}
Heun's equations emerge frequently in modern applications in physics and their importance increases, whereas hypergeometric equations are studied much more and are known in detail. Therefore studying the reductions are important for better understanding Heun's equation and related physical systems. The Heun-to-hypergeometric reductions are important mainly for two reasons: The first reason is that the hypergeometric equations have a vast literature and the second one is that the occurrence of the hypergeometric equations can be related to some symmetries of systems such as SL(2,$\mathbf{R}$) symmetry \cite{cvetbir,Birkandan:2015yda}.

In this paper, we give physical examples to the limiting cases in which the Heun equation reduces to the hypergeometric equation.

We first analyzed the main equation obtained by studying the Coulomb problem on a 3-sphere and reduced the solution which was originally given by the Heun's functions to the hypergeometric functions in some limiting cases. Using the parameter restrictions coming from the reduction cases, we obtained energy conditions which corresponded to the energy spectrum obtained in the original paper.

The second example was the s-wave bound state equation in the problem of the attractive inverse square potential. We found that the nontrivial cases were inapplicable in this problem but the trivial cases simplified the problem. In the physically interesting case with $\Delta =0$ and $q=ab$ we found the same results with a special case studied in the original paper.

The third example was the equation for the limit density function for the discrete-time quantum walk. In this problem, the trivial cases were inapplicable, whereas in the nontrivial case we successfully reduced the equation using a polynomial transformation of degree $3$. We mathematically obtained a limit density function similar to the one that the original paper gave for a special case.

In our last example, charged particle motion under magnetic field and Coulomb force is analyzed and we recovered the limiting case in which the Coulomb repulsion vanishes using the case \{$ab =0$ and $q=0$\}. We also obtained a nontrivial reduction from Heun-to-hypergeometric function.

From the physical implications of the reduction cases, we see that we can reproduce some limiting cases of the physical problems. As a future prospect, however, we need a systematic study of the mathematical correspondence and more physical examples for enabling a concrete understanding of these physical implications of the reduction cases.
\section{Acknowledgments}
The authors would like to thank Prof. M. Horta\c{c}su for fruitful discussions and the organizing committee of the Days on Diffraction 2015. TB would also like to thank Professors A. Ishkhanyan, A. Kazakov, O. V. Motygin and S. Yu. Slavyanov for valuable comments and discussions. We also thank Prof. A. Turbiner for informing us of an error in our reference 9. This work is supported by Istanbul Technical University [\.{I}T\"{U}\ BAP 37519].
\section{Appendix: The Sage code for calculating the identities}
The code given in this appendix can be used to apply the identities mentioned in the text. The code is also accessible via GitHub \cite{github}.
\begin{verbatim}
###########################################################################
# This Sage code (http://www.sagemath.org/)
# is written by Tolga Birkandan as an appendix to the article:
# "Physical Problems Admitting Heun-to-Hypergeometric Reduction",
# arXiv:1401.0449 [math-ph].
#
# See Table 2 of R. S. Maier, "The 192 Solutions of the Heun Equation",
# Math. Comp. 76 (2007), 811-843, arXiv:math/0408317
# (line 5 & line 9 respectively) for the identities.
###########################################################################
reset()
#%%%%%%%%%%%%%%%%%%%%%%%%%%%%%%%%%%%%%%%%%%%%%%%%%%%%%%%%%%%%%%%%%%%%%%%%%%
def iden1(oldvars):
    newvars=[]
    newvars.append(oldvars[0]/(oldvars[0]-1))
    newvars.append((-oldvars[1]+oldvars[2]*oldvars[0]*oldvars[4])/(oldvars[0]-1))
    newvars.append(oldvars[2])
    newvars.append(oldvars[2]-oldvars[5]+1)
    newvars.append(oldvars[4])
    newvars.append(oldvars[2]-oldvars[3]+1)
    newvars.append(oldvars[6]/(oldvars[6]-1).simplify_full())
    writeresult='H( %s, %s; %s, %s, %s, %s; %s) \
    =(1 - %s )^{-%s} H( %s, %s; %s, %s, %s, %s; %s)'%(latex(d1),latex(q1),latex(a1),latex(b1)
                                  ,latex(Gamma1),latex(Delta1),latex(z1),latex(z1), latex(a1)
                                  ,latex(newvars[0]),latex(newvars[1])
                                  ,latex(newvars[2]),latex(newvars[3])
                                  ,latex(newvars[4]),latex(newvars[5]),latex(newvars[6]))
    show(writeresult)
#%%%%%%%%%%%%%%%%%%%%%%%%%%%%%%%%%%%%%%%%%%%%%%%%%%%%%%%%%%%%%%%%%%%%%%%%%%
def iden2(oldvars):
    newvars=[]
    newvars.append(1/oldvars[0])
    newvars.append(oldvars[1]/oldvars[0])
    newvars.append(oldvars[2])
    newvars.append(oldvars[3])
    newvars.append(oldvars[4])
    newvars.append(oldvars[2]+oldvars[3]-oldvars[4]-oldvars[5]+1)
    newvars.append((oldvars[6]/oldvars[0]).simplify_full())
    writeresult='H( %s, %s; %s, %s, %s, %s; %s) \
    =H( %s, %s; %s, %s, %s, %s; %s)'%(latex(d1),latex(q1),latex(a1),latex(b1)
                                  ,latex(Gamma1),latex(Delta1),latex(z1)
                                  ,latex(newvars[0]),latex(newvars[1])
                                  ,latex(newvars[2]),latex(newvars[3])
                                  ,latex(newvars[4]),latex(newvars[5]),latex(newvars[6]))
    show(writeresult)
#%%%%%%%%%%%%%%%%%%%%%%%%%%%%%%%%%%%%%%%%%%%%%%%%%%%%%%%%%%%%%%%%%%%%%%%%%%
# Define the variables:
# The user should define new variables if necessary
d1,q1,a1,b1,Gamma1,Delta1,z1=var('d1,q1,a1,b1,Gamma1,Delta1,z1')
d,q,a,b,Gamma,Delta,z=var('d,q,a,b,Gamma,Delta,z')
#%%%%%%%%%%%%%%%%%%%%%%%%%%%%%%%%%%%%%%%%%%%%%%%%%%%%%%%%%%%%%%%%%%%%%%%%%%
# The user should change this part.
# Enter your parameters here:
d1,q1,a1,b1,Gamma1,Delta1,z1 = d,q,a,b,Gamma,Delta,z
#%%%%%%%%%%%%%%%%%%%%%%%%%%%%%%%%%%%%%%%%%%%%%%%%%%%%%%%%%%%%%%%%%%%%%%%%%%
oldvars=[d1,q1,a1,b1,Gamma1,Delta1,z1]
#%%%%%%%%%%%%%%%%%%%%%%%%%%%%%%%%%%%%%%%%%%%%%%%%%%%%%%%%%%%%%%%%%%%%%%%%%%
# Do the calculation:

# This is how you apply the first identity:
iden1(oldvars)

# This is how you apply the second identity:
iden2(oldvars)
#%%%%%%%%%%%%%%%%%%%%%%%%%%%%%%%%%%%%%%%%%%%%%%%%%%%%%%%%%%%%%%%%%%%%%%%%%%
\end{verbatim}

\end{document}